\documentclass{emulateapj}

\usepackage{apjfonts}

\newcommand{\kms}{\mbox{km s$^{-1}$} } 
\newcommand{\cmthree}{\mbox{cm$^{-3}$ }}

\shortauthors{Narayanan et al.}
\shorttitle{Theoretical Molecular Line Emission from Gravitationally
Unstable Proplyds}

\begin{document}
\title{Molecular Line Emission from Gravitationally Unstable Protoplanetary Disks}

\author{Desika Narayanan\altaffilmark{1,3}, Craig A.
Kulesa\altaffilmark{1}, Alan Boss\altaffilmark{2}, Christopher
K. Walker\altaffilmark{1}}

 \altaffiltext{1}{Steward Observatory,
University of Arizona, 933 N Cherry Ave, Tucson, Az, 85721}
 \altaffiltext{2}{Department of 
Terrestrial Magnetism, Carnegie Institution of Washington, 5241 Broad 
Branch Road, NW, Washington, DC, 20015-1305}
\altaffiltext{3}{dnarayanan@as.arizona.edu}
\slugcomment{Accepted by the Astrophysical Journal}

\begin{abstract}
In the era of high resolution submillimeter interferometers, it will
soon be possible to observe the neutral circumstellar medium directly
involved in gas giant planet (GGP) formation at physical scales
previously unattainable.  In order to explore possible signatures of
gas giant planet formation via disk instabilities, we have combined a
3D, non-local thermodynamic equilibrium (LTE) radiative transfer code
with a 3D, finite differences hydrodynamical code to model molecular
emission lines from the vicinity of a 1.4 M$_{\rm J}$ self-gravitating
proto-GGP. Here, we explore the properties of rotational transitions
of the commonly observed dense gas tracer, HCO$^+$.  Our main results
are the following: 1. Very high lying HCO$^+$ transitions
(e.g. HCO$^+$ J=7-6) can trace dense planet forming clumps around circumstellar
disks. Depending on the molecular abundance, the proto-GGP may be
directly imageable by the Atacama Large Millimeter Array
(ALMA). 2. HCO$^+$ emission lines are heavily self-absorbed through
the proto-GGP's dense molecular core. This signature is nearly
ubiquitous, and only weakly dependent on assumed HCO$^+$
abundances. The self-absorption features are most pronounced at higher
angular resolutions. Dense clumps that are not self-gravitating only
show minor self-absorption features. 3. Line temperatures are highest
through the proto-GGP at all assumed abundances and inclination
angles. Conversely, due to self-absorption in the line, the
velocity-integrated intensity may not be. High angular resolution
interferometers such as the Submillimeter Array (SMA) and ALMA may be
able to differentiate between competing theories of gas giant planet
formation.

\end{abstract}

\section{Introduction}

Since the discovery of 51 Pegasus, there have been numerous detections
of Jupiter-sized extrasolar planets through radial velocity
experiments, transiting of parent stars and direct imaging (for recent
reviews, see Udry, Fischer \& Queloz, 2006, Charbonneau et al. 2006
and Beuzit et al.  2006).

Concomitant to the problem of characterizing the nature of these gas
giant planets (GGPs) is developing the theoretical constructs which
describe the nature of GGP formation. Two major theories have been
developed concerning the physics of GGP formation. Core accretion
begins with the formation of planetesimals through the collisional
coagulation and sticking of progressively larger solid bodies in the
circumstellar environment.  Once $\sim$kilometer sized planetesimals
are formed, runaway accretion to Mars-sized bodies can occur. When the
planetary embryos reach a mass of ten Earth masses or so, disk gas is
accreted dynamically, resulting in GGP formation.  Core accretion is
the generally favored mechanism for forming Jupiter and Saturn
(Pollack et al. 1996; Goldreich, Lithwick, \& Sari, 2004a). The core
accretion model has been challenged in explaining how a solid core can
form on timescales less than the disk dissipation times of
$\sim$10$^6$-10$^7$yr (Pollack et al. 1996; Ikoma, Nakazawa, \& Emori,
2000).  However it should be noted that recent models by Rafikov
(2003) and Goldreich, Lithwick \& Sari (2004b) have suggested a core
accretion mechanism that is not incompatible with the short-lived disk
lifetimes. Additional problems of the core-accretion model include
resolving theoretical GGP core masses with those of Jupiter and Saturn
(Mejia, 2004; Saumon \& Guillot, 2004).

As an alternative to core-accretion, the disk instability mechanism
has been investigated in many models by Boss (2001, 2004), Mejia
(2004), and Mayor et al. (2005). In this theory, marginally unstable
gaseous disks contract gravitationally to form GGPs.  Models of
gravitationally unstable disks suggest planets can form on rapid
($\sim$10$^3$ yr) time scales (Boss 1997, 1998), but require disks to
cool efficiently in order for gravitationally bound clumps to
form. Boss (2004) has shown that convective cooling through
protoplanetary disks can be quite efficient. Hybrid mechanisms for GGP
formation have been suggested as well (Currie, 2005).

It is clear that both leading theories in the formation of GGPs have
their shared successes and challenges in current models. Observations
of protoplanetary disks will be key in constraining the two models.
Because the gravitational instability method involves the accumulation
of large clumps of cool gas from the circumstellar disk, molecular
line observations may be helpful in revealing the nature of GGP
formation. Indeed, in the era of high-resolution millimeter and
submillimeter wave interferometers, clumps of cool and dense gas may
indeed be directly imageable in nearby circumstellar disks.

Rotational transitions ($J$+1$\rightarrow$$J$) in interstellar
molecules have long been used to better understand the nature of cold
gas in star-forming environments (for a recent review, see Evans,
1999). Serving as a proxy for the observationally elusive molecular
hydrogen (H$_2$), excitation analysis of lines from molecules such as
CO, CS, HCO$^+$, HCN and others can provide diagnostics for the
physical conditions in the cold molecular gas.

Through the use of submillimeter and millimeter wave interferometers,
direct imaging of cold circumstellar disks have recently been made
possible (e.g. Patel. et al. 2005, Qi et al. 2004). Submillimeter
molecular line emission has been used to estimate molecular depletion
factors (e.g. Andrews \& Williams, 2005; van Zadelhoff et al., 2001),
as well as provide information concerning kinematics in the
circumstellar environment (e.g. Qi et al. 2003). It is additionally
possible to constrain disk molecular gas masses through such
observations (e.g. Hogerheijde et al. 2002). Recent CO (J=6-5)
observations of TW Hya by Qi et al. (2006) has shown that molecular
excitation can be quite high due to the dense gas present in
protoplanetary disks.

\begin{figure}
\vspace{0.7cm}
\includegraphics[angle=90,scale=0.35]{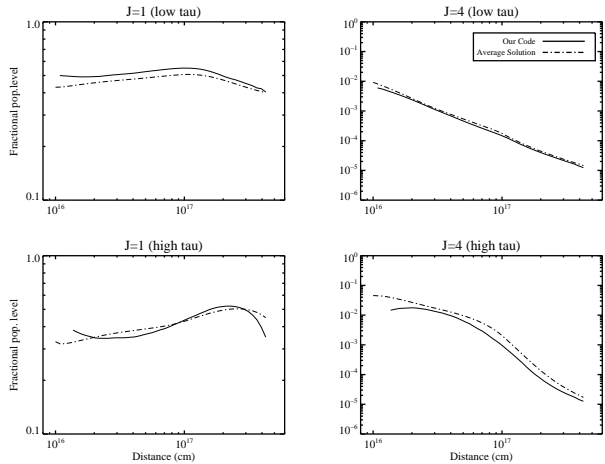}
\caption{Solutions of our radiative transfer codes to test problems
  ``2a'' and ``2b'' of van Zadelhoff et al. (2002). The ordinate is
  fractional populations of the J=1 and J=4 levels of HCO$^+$ and the
  abscissa is radial distance out from center of a 1-dimensional
  inside-out collapsing cloud. The solid line is our solution, and the
  dash-dot line is the average solution to the problem taken from
  seven other researcher's codes.\label{figure:solution}}
\vspace{0.5cm}
\end{figure}

In addition to observational data sets of molecular line emission from
protoplanetary disks, radiative transfer modeling of the emission 
patterns can provide powerful constraints as to the physical
conditions in the molecular gas.  As an example, Qi et al. in the
aforementioned study of TW Hya used a 2D non-LTE Monte Carlo model to
derive vertical temperature distributions in the disk. Similar studies
have been performed to calculate temperature and/or density
distributions by Kessler (2004), and Semenov et al. (2005), among many
others.

In order to fully utilize the new generation of interferometers to
probe planet formation, submillimeter and millimeter wave emission
modeling of GGP forming disks is needed. Progressive works by Wolf \&
D'Angelo (2005), Moro-Martin, Wolf \& Malhotra (2005), and Varniere et
al.  (2006) have studied the effects of protoplanets in disks on
infrared SEDs through the use of radiative transfer modeling. A
necessary complementary dataset to these works is models of molecular
line emission from planet forming circumstellar disks.

In this study, we take the first step in this direction by
applying a newly developed 3D non-LTE radiative transfer code to a
model of a gravitationally unstable protoplanetary disk (Boss, 2001).
We will discuss the emission patterns by way of contour maps and
emission line profiles. This paper is organized as follows: in \S2, we
discuss the numerical methods involved concerning hydrodynamics,
radiative transfer, and chemistry; in \S3 we present synthetic images
of HCO$^+$ emission in the disk; in \S4 we discuss emission line
profiles; in \S5 we briefly discuss spectral maps and in \S6 we
summarize.

\section{Numerical Methods}
\subsection{Hydrodynamics}
\label{section:hydro}

\begin{figure}
\includegraphics[angle=270,scale=0.4]{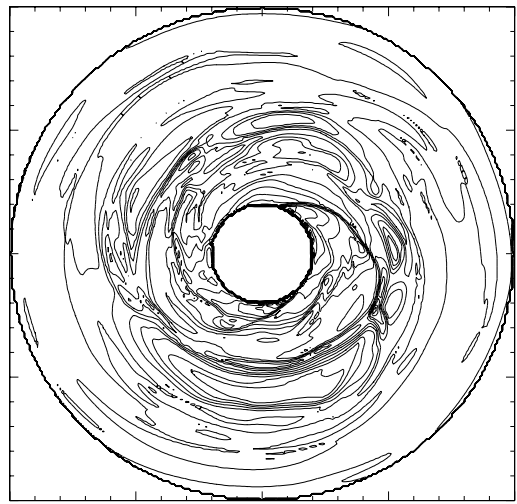}
\caption{Equatorial density contours for hydrodynamic
snapshot. Contours denote changes by a factor of 2 in density, with
$\rho_{\rm max}$=5.0$\times$10$^{-9}$gm cm$^{-3}$. The maximum density
clump is located at $\sim$4 o'clock in this image.\label{figure:denscontours}}
\vspace{1cm}
\end{figure}

\begin{figure*}
\includegraphics[angle=90,scale=0.9]{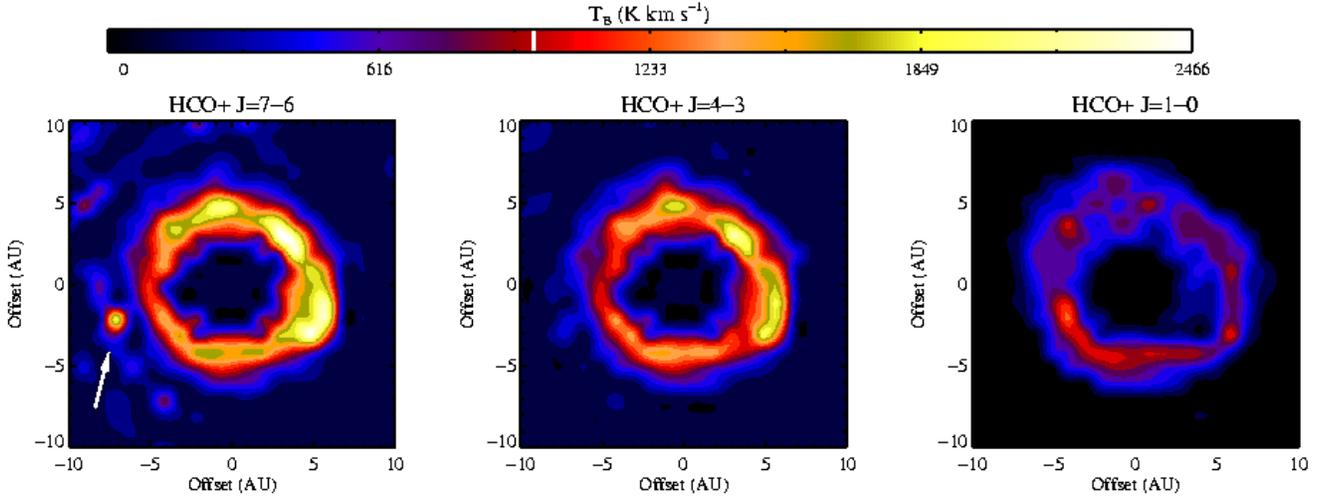}
\caption{The effect of transition choice on the image: HCO$^+$
transitions J=7-6, 4-3, 1-0. The proto-GGP is evident only in the
highest HCO$^+$ transitions due to the high critical density needed to
image the dense object. The synthesized images are the mirror image of
the density contours in Figure~\ref{figure:denscontours}, and thus the
proto-GGP is at $\sim$8 o'clock in the left most panel. The intensity
is in units of K-\kms and is on a fixed scale for the entire
figure.\label{figure:map1}}
\end{figure*}

We have run three dimensional hydrodynamic models using a
finite-differences code to model the gravitationally unstable
protoplanetary disk as fully described in Boss (2001). The code has
been shown to be accurate to second order in space and time (Boss \&
Myhill 1992). The spherical grid is uniformly spaced in the radial
direction between 4 and 20 AU.  The azimuthal grid is uniform, and the
polar grid spaced such that the resolution grows toward the midplane
(with a maximum resolution of $\Delta \theta$=0.3$\degr$). The model
includes a central protostar of 1 M$_\sun$ and a disk of 0.091
M$_\sun$. The protostar wobbles in response to the evolving disk, such
that the center of mass of the system is preserved. The initial disk
surface density is assumed to vary with radius as $\sigma \propto
r^{-1/2}$ to $\sigma \propto r^{-1}$ through the inner disk, and
$\sigma \propto r^{-3/2}$ in the outer disk. The initial density
distribution is seeded with perturbations of the form cos($m\phi$)
where $m$=1,2,3,4 with an amplitude of 0.01. Random noise is included at
a lower amplitude.

\subsection{Radiative Transfer}
We build the emergent spectrum by integrating the equation of
radiative transfer over numerous lines of sight through the
hydrodynamic snapshot (e.g. Walker, Narayanan \& Boss 1994). The
solution to the equation of radiative transfer has the numerical form
\begin{equation}
  I_\nu = \sum_{z_0}^{z}S_\nu (z) \left [ 1-e^{-\tau_\nu(z)} \right
  ]e^{-\tau_\nu(\rm tot)}
\end{equation}
where $S_\nu$ is the source function, and $\tau$
is the optical depth.  If local thermodynamic equilibrium (LTE)
conditions hold, then the source function can be replaced by the
Planck function.  LTE is a fair approximation when the density is much
greater than the critical density and collisions thermalize the gas,
e.g.
\begin{equation}
n \gg \frac{A_{\rm ul}}{ \sum C_{\rm ul}}
\end{equation}
where $A$ is the Einstein rate coefficient for spontaneous emission,
$C$ are the collisional rate coefficients, and $u$,$l$ correspond to
the upper and lower states of a given transition. However, when
considering either low density environments, or molecular species with
high dipole moments (and thus high Einstein $A$ coefficients), the
approximation of LTE may no longer hold.  In these situations, the
source function must be calculated explicitly.

In order to find the source function, an iterative procedure can be
employed. We have developed a three-dimensional non-LTE radiative
transfer code based on the Monte Carlo method.  The first work to
detail a Monte Carlo algorithm for non-LTE line transfer was by Bernes
(1979). Future workers improved the algorithms and expanded to two and
three dimensions (Choi et al. 1995, Juvela et al. 1997 Park \& Hong
1998, Hogerheijde \& van der Tak 2000, Schoier, 2000).  Monte Carlo
techniques in radiative transfer are powerful in that they offer a
large amount of flexibility for different geometries and are easily
parallelized. Due to the statistical nature of Monte Carlo, the
solutions include shot noise, and can be slow to converge in cases of
high ($\tau \gtrsim$100) optical depths.

The non-LTE radiative transfer code we have developed operates on the
following principles, adapted from Bernes (1979): the goal is to solve
for the steady state distribution of energy states among the
molecules.  Once these level populations are known, the source
function for a given transition can be calculated by:
\begin{equation}
S_{\nu}=\frac{n_{u}A_{ul}}{(n_lB_{lu}-n_uB_{ul})}
\end{equation}
where $B$ are the Einstein rate coefficients for absorption and
stimulated emission. However, the problem is circular: the level
populations in any grid cell depend on the mean intensity field
through that point.
\begin{equation}
J_\nu=\frac{1}{4\pi}\int I_\nu d\Omega
\end{equation}
The intensity field depends on the emission from other cells which
in turn is given by their source functions.  Hence, it is necessary to
guess the level populations, solve for the mean intensity field,
calculate updated level populations and iterate until the level
populations have converged.

The radiation field is modeled by photon 'packets' that represent many
real photons. The number of photons each model packet represents is
proportional to the Einstein-$A$ for the transition and the number of
molecules or atoms in the upper state of the transition in the
emitting grid cell. The photons are isotropically emitted in a
spontaneous manner with a line frequency drawn from the line profile
function:
\begin{equation}
\phi(\nu)=\frac{1}{\sigma \sqrt{\pi}}{\rm exp}\left \{-\left
  (\nu-\nu_0-\bf{v \cdot \hat n} \frac{\nu_{ul}}{c}\right )^2 /
  \sigma^2\right \}
\end{equation}
where $\nu$ is the frequency of the emitted photon, $\nu_0$ is the
rest frequency of the transition, ${\bf v}$ is the velocity of the
emitting clump of gas, and $\sigma$ is the standard deviation of the
profile function.  The standard deviation is the Doppler width
determined by the local kinetic temperature and microturbulent velocity:
\begin{equation}
\sigma=\frac{\nu_0}{c}\left [\frac{2kT}{m}+V_{\rm turb}\right ]^{\frac{1}{2}}
\end{equation}
We assume a constant microturbulent velocity of 0.2 \kms. The
photon then takes a step of a given distance $s$, passing through gas
with opacity:
\begin{eqnarray}
\alpha_\nu(\rm dust)=\kappa_\nu \rho_{\rm dust}\\ 
\alpha_\nu^{ul}(\rm
gas)= \frac{h \nu_{ul}}{4\pi}\phi(\nu)(\it n_lB_{lu}-n_uB_{ul})
\end{eqnarray}

After passing through this grid cell, the number of real photons the
model photon packet represents is diminished by a factor $e^{-\tau}$
due to absorptions. The model photon continues to take steps in the
same direction until it either leaves the grid or the number of
photons it represents has become negligible. When
all of the photons have been emitted, the mean intensity is known
through each grid point.

The updated level populations are then calculated with the equations of 
statistical equilibrium 
\begin{eqnarray*}
  n_l\left [\sum_{k<l}A_{lk}+\sum_{k\neq l}(B_{lk}J_{\nu}+C_{lk})\right]= \\
  \sum_{k>l}n_kA_{kl}+\sum_{k\neq l}n_k(B_{kl}J_\nu+C_{kl})
\label{stateq}
\end{eqnarray*}
which are solved through standard matrix inversion methods. With the
new level populations in hand, the process can be repeated with a new
calculated radiation field. These radiative transfer calculations are
then iterated over until the level populations are found to converge.
Convergence in non-LTE radiative transfer simulations depends both on
the number of model photons realized in the iteration, as well as the
optical depth.  

The level population calculations are sensitively dependent on the
accuracy of the rate coefficients. We have obtained our coefficients
from the {\it Leiden Atomic and Molecular Database}\footnote{
http://www.strw.leidenuniv.nl/$\sim$moldata} (Schoier et al. 2005). In order
to independently test our radiative transfer codes, we have run the
test problem of an inside-out collapsing sphere published by van
Zadelhoff et al. (2002). We present the results of this test in Figure
\ref{figure:solution}, and direct the reader to van Zadelhoff et al. (2002)
for details on the test problem and solutions.

\subsection{The Model and Parameters}

We have run our non-LTE radiative transfer code for HCO$^+$ rotational
transitions through model HR of Boss (2001). In this model, a 1.4
Jupiter-mass dense clump of gas is formed through gravitational
instabilities in the circumstellar disk. The maximum density through
the proto-GGP is 5.0$\times$10$^{-9}$g cm$^{-3}$, and it orbits at a
semimajor axis of $\sim$10 AU. The temperature through the proto-GGP 
ranges from $\sim$100-150 K.

 For the radiative transfer calculations, we considered HCO$^+$
molecular line emission, as well as thermal continuum emission with
opacities given by Boss \& Yorke (1990). The radiative transfer
calculations were run on one snapshot of the hydrodynamic model for
which we show the midplane density contours in
Figure~\ref{figure:denscontours}. The boundary conditions for the
non-LTE transfer included photons from the 2.73 K cosmic microwave
background. However, when convolving our resultant images to beams
larger than the grid, we assumed a vacuum boundary. Our grid for the
calculations was spherical in nature with
($N_r$,$N_\phi$,$N_\theta$)=(100,512,43). We emitted roughly
3.11$\times$10$^8$ photons per iteration. The calculations typically
took 2 weeks on 14, 2GHz, AMD-64 processors.

\subsection{Chemistry of HCO$^+$}
\label{section:chemistry}

Abundances in protoplanetary disks can be wildly variable due to the
wide range of temperatures and densities involved, in addition to
variable X-ray and UV fluxes. Our purpose is to investigate gross
emission features from dense gas clumps in circumstellar disks and a
full model of the complex chemical reaction networks in disk
environments (e.g. Aikawa \& Herbst, 1999; Semenov, Wiebe, \& Henning
2004) is beyond the scope of this paper. We approximate the impact of
disk chemistry by running models at different HCO$^+$ abundances and
analyzing their imprint on emission features.

 There are commonly two methods of introducing chemical depletion in
disks: a uniform depletion factor across the entire disk, and a jump
depletion where grain depletion only comes into play below a certain
temperature threshold. The formation of HCO$^+$ depends directly on
the abundance of CO, and thus the HCO$^+$ abundance is assumed to
follow the CO characteristic depletion.  CO freeze-out typically
occurs at T$\leq$22K, thus implying a necessary HCO$^+$ depletion at
low temperatures.  However, the abundance of HCO$^+$ can be altered
from typical interstellar values for a variety of other factors:
photodissociation can occur in the hot inner regions of the disk due
to the enhanced UV flux of a young star, or through interstellar
cosmic rays.  We therefore utilize a uniform depletion from standard
ISM abundances taken from Lee, Bettens \& Herbst, (1996).  This
assumption may limit the predicted detectability of massive gas clumps
in a circumstellar disk. In a study of the fractional ionization in
disks, Semenov et al. (2004) find the dominant ion in the intermediate
layer and midplane to be HCO$^+$ at the radii where the densest gas
clumps in our models form (10 AU), whereas the fractional HCO$^+$
drops in the surface layers. By uniformly depleting the HCO$^+$
abundance, we may be decreasing emergent flux.

 van Zadelhoff et al. (2001) estimate depletions ranging from
 0.1-0.001 from standard ISM abundances for disks TW Hya and LkCa 15.
 Dutrey et al. (1997) find an average HCO$^+$ abundance of
 7.4$\times$10$^{-10}/$H$_2$, corresponding to a depletion of
 $\sim$0.15. We have run three models corresponding to depletion
 factors of 0.5, 0.1 and 0.01. While simple depletion models are
 likely not valid for the extremely dense cores of proto-GGPs, as we
 discuss in \S~\ref{section:pumping}, the specifics of the chemistry
 in the densest regions of the disk may not contribute significantly
 to the emergent HCO$^+$ flux.

%The results we derive should be interpreted as those for dense gas
%clumps directly involved in GGP formation. The results do not trace an
%evolved protoplanet itself, where complex chemical reaction networks
%almost certainly disallow simple depletion approximations.

\section{Images}

\begin{figure}
%bb=54 54 458 738
\includegraphics[angle=90,scale=0.425]{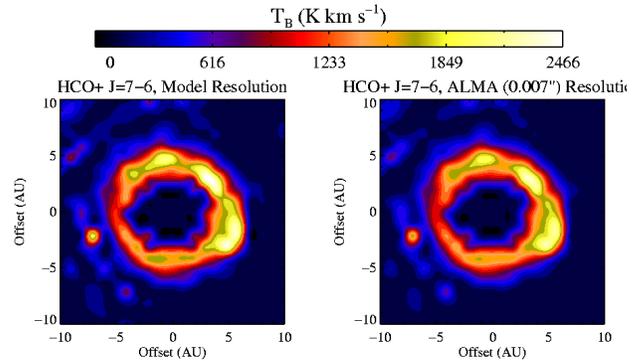}
\caption{The effect of angular resolution on the image: HCO$^+$
transitions J=7-6. The left column is at model resolution. The right
column is a simulated ALMA image at 0.007 $\arcsec$ (the most extended
baseline of ALMA, Bastian, 2002) for a source distance of 140 pc. The
synthesized images are the mirror image of the density contours in
Figure~\ref{figure:denscontours}, and thus the proto-GGP is at $\sim$8
o'clock in each panel. The intensity is in units of K-\kms and is on a
fixed scale for the entire figure.\label{figure:map2}}
\end{figure}

High angular resolution interferometers such as ALMA will be able to
achieve unprecedented imaging capabilities of Galactic circumstellar
disks. Utilizing our radiative transfer codes, we have created
synthetic intensity contour maps of the HCO$^+$ emission from the
gravitationally unstable protoplanetary disk in our models, and
present them in this section.  The parameter space we explore 
includes rotational transition, inclination angle, and abundance. 

\subsection{Molecular Transition}
%,bb=0 54 300 738
\begin{figure*}
\includegraphics[angle=90,scale=0.9]{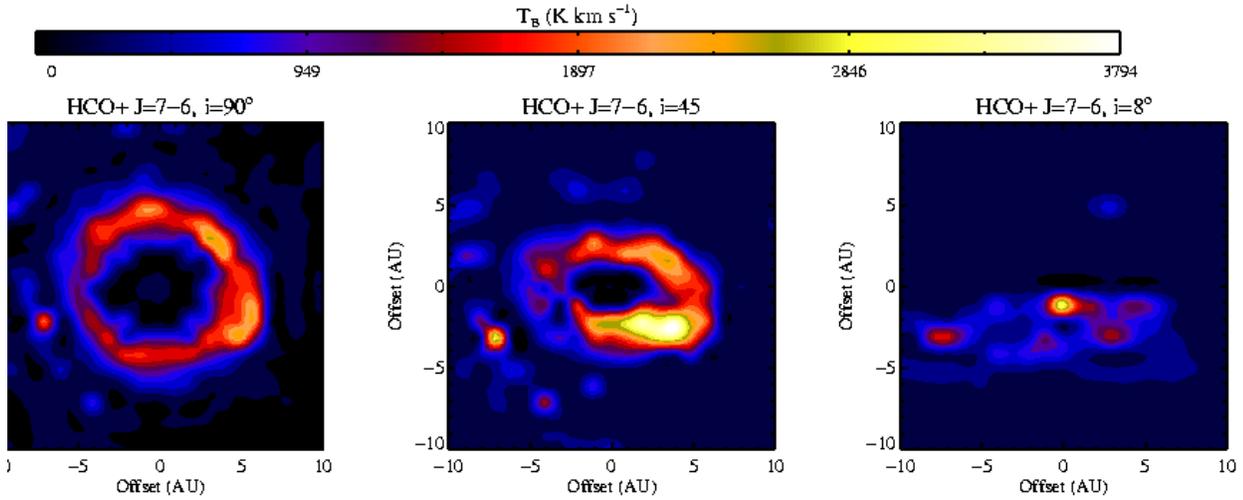}
\caption{The effect of inclination on the image: HCO$^+$ transition
J=7-6. The inclinations are 90$\degr$ (face on), 45$\degr$ and
8$\degr$. 8$\degr$ is roughly the minimum inclination angle at which
the proto-GGP did not get lost in the emission of the disk. The
synthesized images are the mirror image of the density contours in
Figure~\ref{figure:denscontours}, and thus the proto-GGP is at $\sim$8
o'clock in each panel.\label{figure:map3}}
\end{figure*}
%,bb=0 54 375 738
\begin{figure*}
\includegraphics[angle=90,scale=0.9]{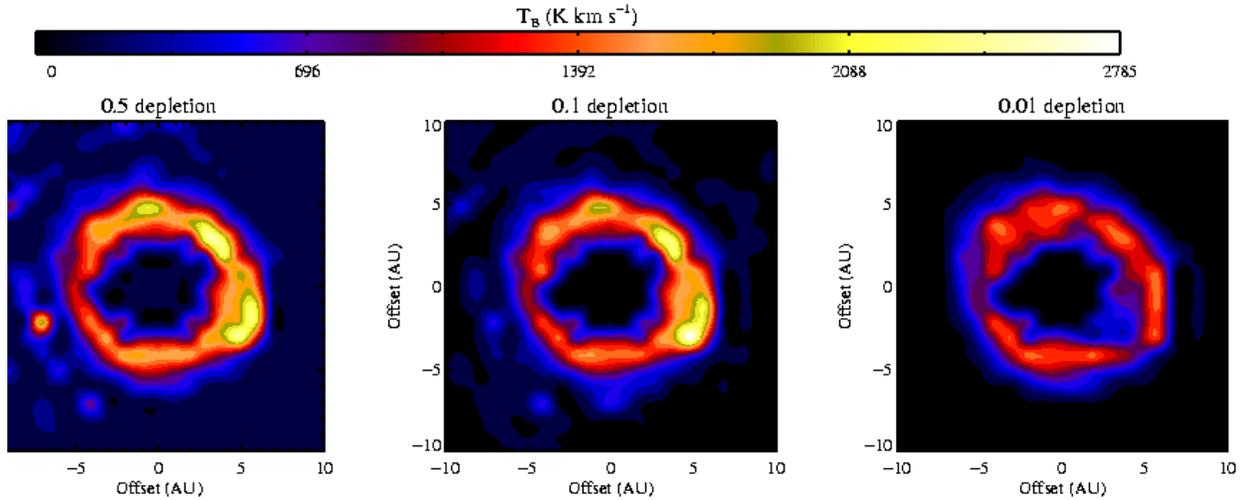}
\caption{The effect of HCO$^+$ abundance on the
 image. Abundances (from left) are 1$\times$10$^{-9}$/H$_2$,
 5$\times$10$^{-10}$/H$_2$, 5$\times$10$^{-11}$/H$_2$. The proto-GGP
 is only visible in the image at highest HCO$^+$ abundance. The
 synthesized images are the mirror image of the density contours in
 Figure~\ref{figure:denscontours}, and thus the proto-GGP is at
 $\sim$8 o'clock in the leftmost panel.\label{figure:map4}}
\end{figure*}

In Figure \ref{figure:map1} we show model contours of spectral line
intensity of the protoplanetary disk as viewed through the HCO$^+$ J=1-0,
4-3 and 7-6 transitions at 1/2 ISM abundance. The disk is face on
($i$=90$\degr$) in each image.

The GGP in our simulation achieves a maximum density of
$n$=3.0$\times$10$^{15}$cm$^{-3}$. At these high densities, and warm
temperatures ($\sim$100 K), collisions will ensure that most of the
HCO$^+$ molecules are excited well above the ground state. Indeed, as
our simulations show, the GGP is only fully visible at the highest
HCO$^+$ transitions. While lower transitions in HCO$^+$ may be able to
detect dense clumps in the circumstellar disk, the densest clumps that
may be self-gravitating can be identified using high-density
tracers. For the HCO$^+$ J=7-6 transition ($n_{\rm crit}\sim$10$^7$
cm$^{-3}$), the proto-GGP emits quite prodigiously while the clumps in
the remaining parts of the disk begin to fade. The high J levels of
dense gas tracers serve as an efficient method of filtering out low
density gas that may not be directly associated with the
protoplanet. We have not explored any transitions beyond HCO$^+$ J=7-6. It may
be, however, that the proto-GGP is even more distinct at higher (THz)
HCO$^+$ transitions. For the transitions we have modeled, the densest
clumps in the circumstellar disk are most visible in HCO$^+$ J=7-6. We
will explore the effects of inclination and abundance primarily in
this transition.

With its most extended baseline, ALMA will be able to achieve a
spatial resolution of 0.007$\arcsec$ at HCO$^+$ (J=7-6, $\nu_0$=624
GHz) (Bastian 2002). At the distance of Taurus Molecular Cloud,
$\sim$140pc, this angular resolution (1 AU) is quite comparable to our
model resolution of $\sim$0.5 AU.  In Figure \ref{figure:map2}, we
have simulated an observation of our circumstellar disk by setting it
at a distance of 140 pc and convolving it with the 0.007$\arcsec$ ALMA
beam. We assume no lost flux and a circular Gaussian beam. The
proto-GGP and other dense clumps are quite visible. As an example, in
one hour of integration, the 64-element ALMA array should be able to
image the dense protoplanet in Figure \ref{figure:map2} with a signal
to noise ratio of $\sim$3-4.  Additionally, as we will show in
\S~\ref{section:lineprofiles}, signatures of dense clumps may be
evident even in single-dish sub-mm telescopes through signatures in
the emission line profile.

\begin{figure}
\includegraphics[angle=90,scale=0.36]{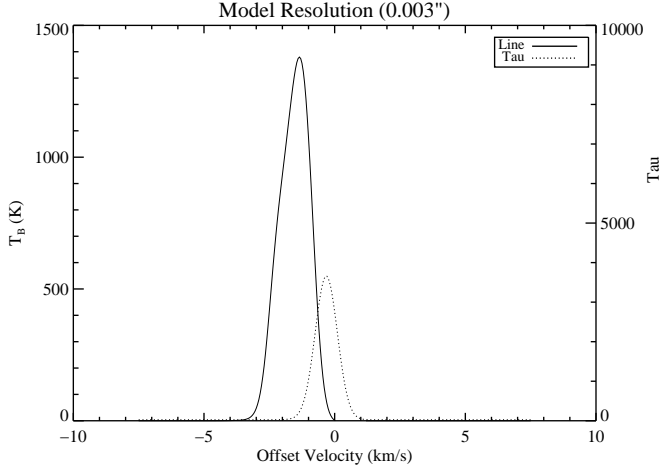}
\caption{HCO$^+$ J=7-6 spectral line profile at
model resolution through the proto-GGP with optical depth
overplotted. The central density is $\sim$10$^{15}$\cmthree, and the
line is completely self-absorbed at line center. The emission is
primarily due to radiatively pumped gas in the outer layer of the
proto-GGP. \label{figure:ggp_spectra}}
\end{figure}

\begin{figure}
\vspace*{1cm}
\includegraphics[angle=90,scale=0.36]{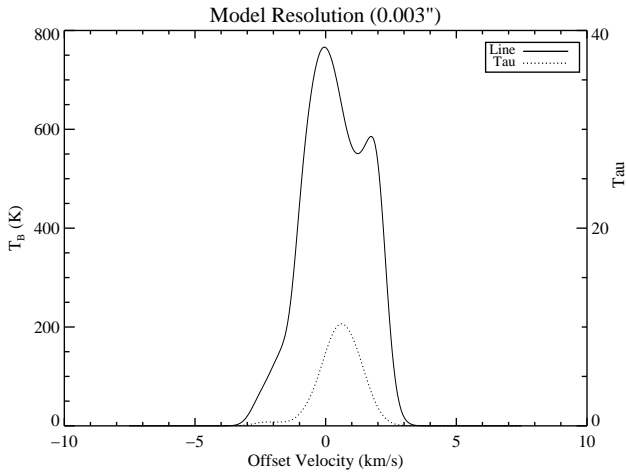}
\caption{HCO$^+$ J=7-6 spectral line profile at
model resolution through the dense clump at position x = 5 AU,
y = -3 AU (for reference with Figure~\ref{figure:map4}).  The central
density of the clump is $\sim$10$^{15}$\cmthree. Some emission from
the central regions escapes, and the line is not completely
self-absorbed.\label{figure:clump_spectra}}
\vspace{0.25cm}
\end{figure}

\subsection{Inclination}
 As we will discuss in \S~\ref{section:pumping}, the emission from the
proto-GGP is largely from the outer layers. Consequently, the
inclination angle of the disk does not affect the results
significantly. As seen in Figure \ref{figure:map3}, even at nearly
edge-on inclinations, the proto-GGP is still visible at HCO$^+$
J=7-6. However, due to increasing column through the disk in
low-inclination scenarios, other dense clumps in the disk begin to
emit at similar intensities as the proto-GGP, thus confusing
observations. Because of this effect, we have found that the minimum
angle in our models that the proto-GGP is visible is $i\sim$8$\degr$.
Inclination effects can quickly wash out signatures of the proto-GGP
in the spectral line profile (see \S\ref{section:lineprofiles}).

\begin{figure}
\includegraphics[angle=90,scale=0.35]{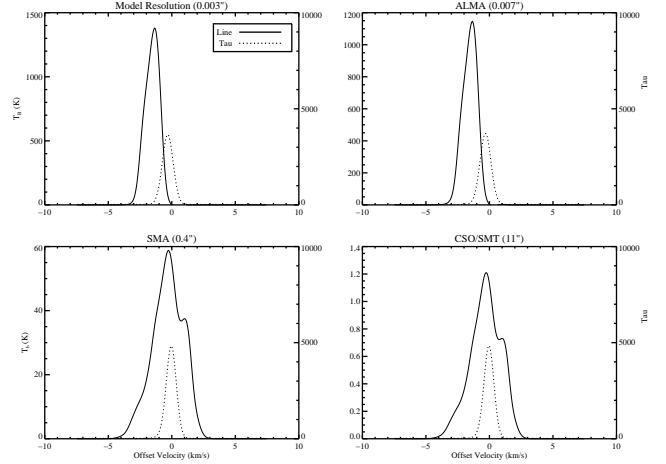}
\caption{HCO$^+$ J=7-6 spectral line profile of proto-GGP,
convolved to the beam of different telescopes. The assumed abundance
is 1$\times$10$^{-9}$/H$_2$ (0.5 depletion from standard ISM values,
Lee, Bettens \& Herbst, 1996). In all cases, the simulated telescope
is pointed directly at the proto-GGP. The telescopes simulated are
ALMA, the SMA and a 10m class single dish. The corresponding optical
depth, $\tau$, is overplotted, with values on the right side of each
box. The disk is face-on.\label{figure:1e-9spec}}
\vspace{0.25cm}
\end{figure}

\subsection{Abundances}
We have modeled the HCO$^+$ abundances as uniform, modulo depletion
factors (\S ~\ref{section:chemistry}). In Figure \ref{figure:map4}, we
present face-on images of the protoplanetary disk at 0.5, 0.1 and 0.01
depletion factors from standard ISM abundances. The proto-GGP is no
longer visible at lower abundances. This holds for all inclinations,
as well as all lower HCO$^+$ transitions. With less emitting molecules
along the line of sight, the proto-GGP emission is no longer able to
dominate over the other dense clumps in the disk. However, the
proto-GGP does, in fact, emit similar peak line temperatures as other
dense clumps in the cloud. Why, then, does the protoplanet not appear
to emit brightly in the image? The image shows velocity integrated
line intensity - the total area under the emission spectrum. Much of
the emission from the proto-GGP is self absorbed, and the majority of
the emission comes from radiatively pumped gas in the outer layers at
low velocity dispersion (\S~\ref{section:pumping}). This effect
reduces the velocity integrated intensity.

\begin{figure}
\includegraphics[angle=90,scale=0.35]{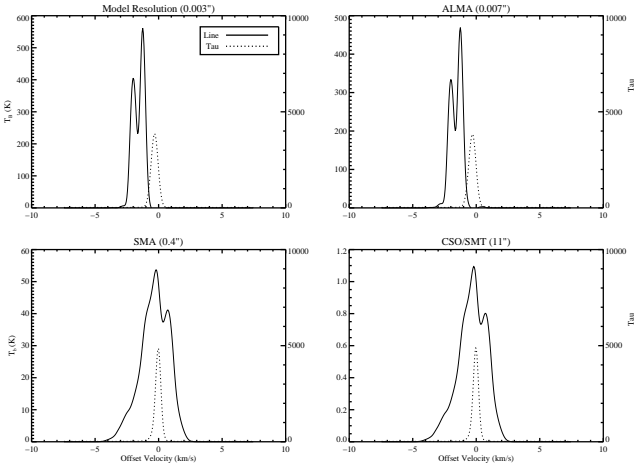}
\caption{HCO$^+$ J=7-6 spectral line profile of proto-GGP,
convolved to the beam of different telescopes. The assumed abundance
is 5$\times$10$^{-10}$/H$_2$ (0.1 depletion from standard ISM values,
Lee, Bettens \& Herbst, 1996). In all cases, the simulated telescope
is pointed directly at the proto-GGP. The telescopes simulated are
ALMA, the SMA and a 10m class single dish. The corresponding optical
depth, $\tau$, is overplotted, with values on the right side of each
box. The disk is face-on.\label{figure:5e-10spec}}
\end{figure}

\section{Line Profiles}
\label{section:lineprofiles}

\subsection{Non-LTE Effects: Radiative Pumping in the Vicinity 
  of Dense Gas Clumps}
\label{section:pumping}

In Figure~\ref{figure:ggp_spectra}, we have plotted the HCO$^+$
(J=7-6) emission spectrum through the proto-GGP in the model with 0.5
depletion factor. The large optical depths cause the line flux to be
$\sim$0 K at line center. The emission from the proto-GGP does not
originate at the core of the dense clump of gas, but rather in the
more diffuse outer layers. While the gas in the dense core of the
proto-GGP is in LTE, the emission from this gas suffers heavy
extinction owing to the high optical depths at line center. However,
the density through the proto-GGP drops off quickly with radius
allowing gas in the outer layers (where $\tau \la$1) to be radiatively
pumped by $\sim$1500 K gas near the protostar. The strong emission
line temperatures, then, originate in radiative excitation of lower
density non-LTE gas, rather than from the core of the cold proto-GGP,
and consequently does not reflect the kinetic temperature of the
emitting gas.

 The emission pattern of heavy self-absorption at line center, and
radiatively pumped gas at the surface is characteristic only of the
densest clump of gas in our models, e.g. the self-gravitating
proto-GGP with central density $\sim$10$^{15}$ \cmthree.  As it is
these densest clumps that serve as antecedents to GGPs, the line
profile of heavy self absorption at line center combined with offset
emission may be characteristic of GGPs in formation.  The optical
depth in other dense clumps in the circumstellar disk is sufficient to
produce self-absorption in the line profiles, but not to the same
degree as seen toward the proto-GGP. As an example, in Figure
~\ref{figure:clump_spectra}, we plot the emergent HCO$^+$ J=7-6
spectra at model resolution and depletion factor 0.5 from a clump in
the face-on disk (located at x = 5 AU, y = -3 AU for reference with
Figure~\ref{figure:map4}) with central density
$\sim$1.75$\times$10$^{11}$\cmthree. The optical depth at line center
only reaches $\tau \approx 1$ near the dense core of the clump, 
allowing significant emission from both the LTE core, as well as the
radiatively pumped outer layers of the dense clump to escape. In
contrast, the optical depth at line center at the core of the
proto-GGP reaches a total value of several thousand, rendering the
majority of the proto-GGP optically thick.

The nature of emission from the proto-GGP has implications concerning
the assumed chemistry in these models.  High optical depths through
the densest regions of the proto-GGP prevents radiation from emerging.
For example, even in the model with lowest optical depths (0.01
depletion factor), only $\sim$5\% of the emission at the line peak
originating from the center of the proto-GGP reaches the observer. This
pales in comparison to the emission from the radiatively pumped gas in
the outer layers of the proto-GGP which has an order of magnitude
higher source function than the LTE gas at the core. The emission due
to radiative pumping of the less dense gas in the outer layers of the
proto-GGP will tend to mask more complex chemistry that may be
occurring in the denser regions.

\begin{figure}
\includegraphics[angle=90,scale=0.35]{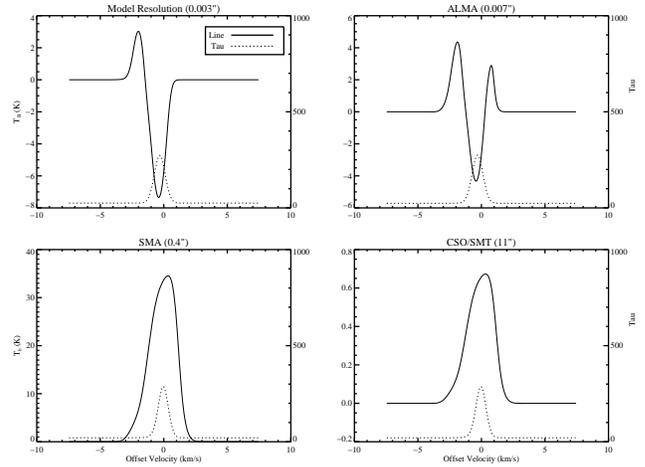}
\caption{HCO$^+$ J=7-6 spectral line profile of proto-GGP,
convolved to the beam of different telescopes. The assumed abundance
is 5$\times$10$^{-11}$/H$_2$ (0.01 depletion from standard ISM values,
Lee, Bettens \& Herbst, 1996). In all cases, the simulated telescope
is pointed directly at the proto-GGP. The telescopes simulated are
ALMA, the SMA and the 10m class single dish. The corresponding optical
depth, $\tau$, is overplotted, with values on the right side of each
box. The disk is face-on.\label{figure:5e-11spec}}
\end{figure}

\subsection{Effects of Abundance and Resolution}
In order to further quantify the effect dense clumps have on the
emission line profiles, we have convolved the model results with
circular Gaussian telescope beams of different sizes (corresponding to
the diameter of the telescope and the frequency of interest) and
plotted the synthesized emission line spectra for three different
HCO$^+$ abundances.

\begin{figure*}
\includegraphics[angle=90,scale=.7]{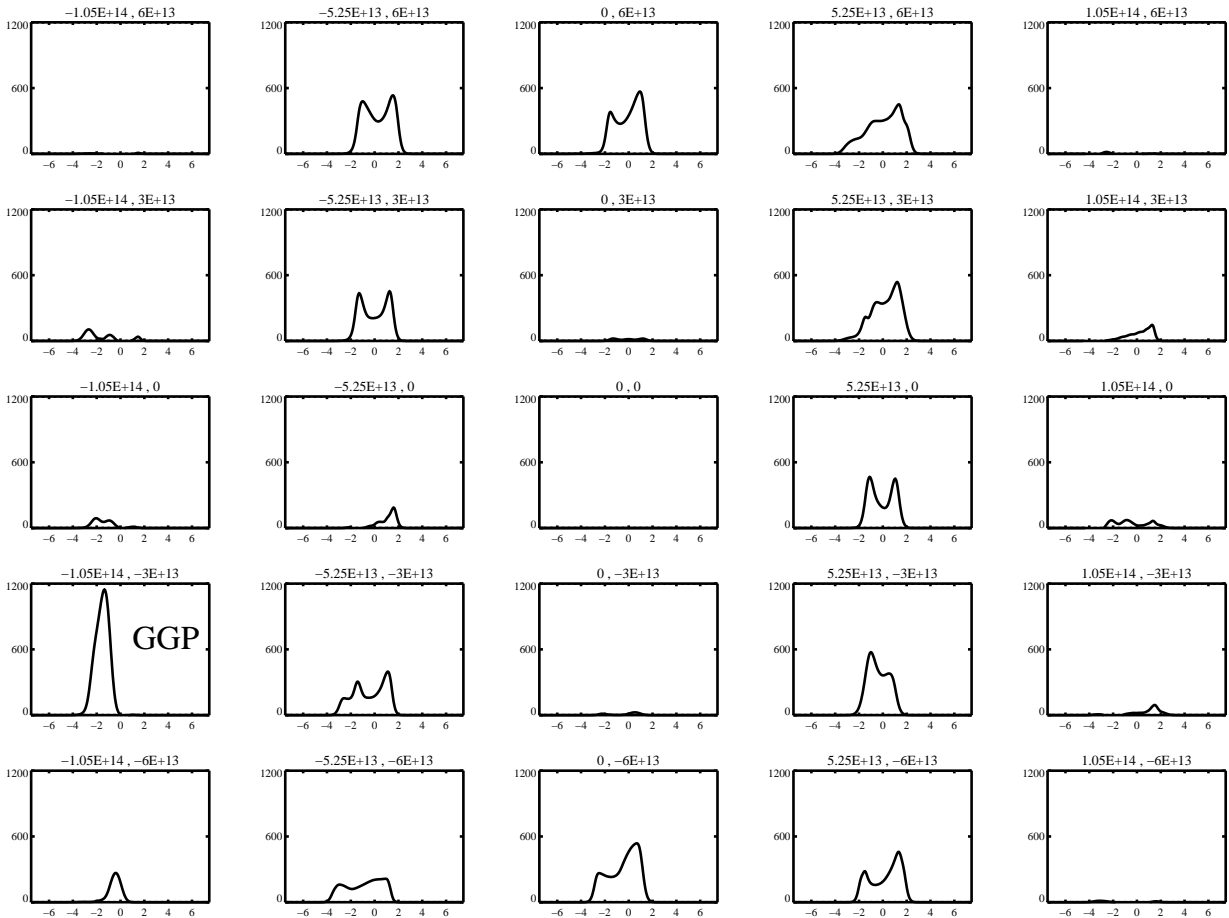}
\caption{Spectral map of model with
1$\times$10$^{-9}$/H$_2$ abundance.  The disk is face on, and the beam
is 0.007$\arcsec$, corresponding to the longest ALMA baseline. The
x-axis in each plot is offset velocity (\kms) and the y-axis in
Rayleigh-Jeans Temperature (K). The title in each panel refers to the
offset position in cm. The map is made from the mirror image of the
density contours presented in
Figure~\ref{figure:denscontours}.\label{figure:1e-9specmap}}
\end{figure*}

\begin{figure*}
\includegraphics[angle=90,scale=.7]{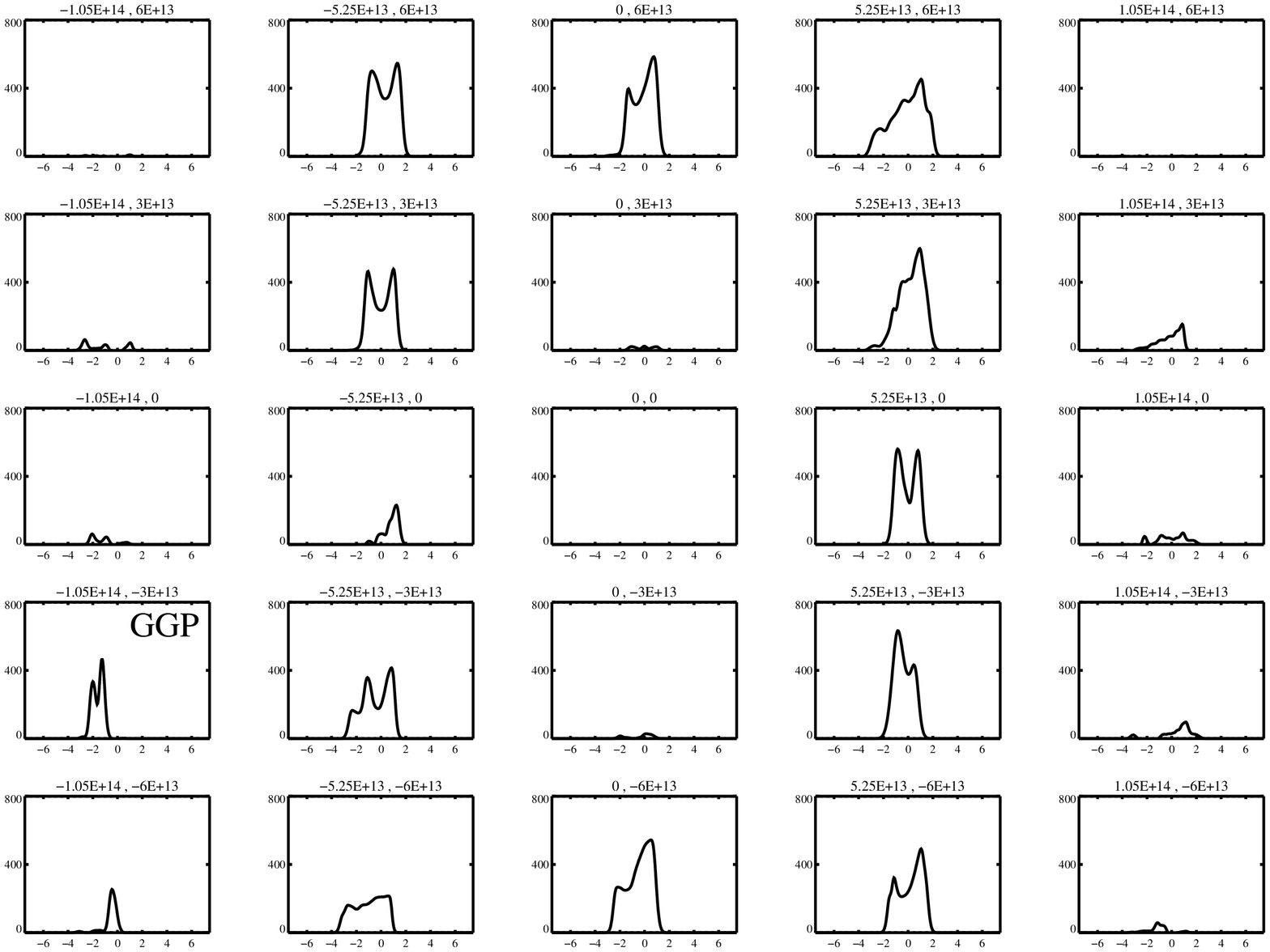}
\caption{Spectral map of model with
5$\times$10$^{-10}$/H$_2$ abundance.  The disk is face on, and the
beam is 0.007$\arcsec$, corresponding to the longest ALMA
baseline. The x-axis in each plot is offset velocity (\kms) and the
y-axis in Rayleigh-Jeans Temperature (K). The title in each panel
refers to the offset position in cm. The map is made from the mirror
image of the density contours presented in
Figure~\ref{figure:denscontours}.\label{figure:5e-10specmap}}
\end{figure*}

In Figure \ref{figure:1e-9spec}, \ref{figure:5e-10spec} and
\ref{figure:5e-11spec}, we have plotted the model emission line
profile as viewed by ALMA, the SMA and 10m class single-dish
telescopes [e.g. Caltech Submillimeter Observatory (CSO) or the
Heinrich Hertz Submillimeter Telescope (SMT)]. With each spectral
line, we have overplotted the optical depth ($\tau$) as a function of
velocity. The model disks are viewed face on. The line profiles are
found to change significantly as a function of abundance. However, in
each instance the line profiles through the proto-GGP are heavily
self-absorbed.

In the model with a 0.5 depletion factor, the emission line through
the proto-GGP, both at model resolution and that of ALMA, is
self-absorbed to zero flux at line center. This is evident from the
location of the peak in the optical depth profile.  The asymmetry in
the line profile ($\sim$1-2 \kms of line center) is due to motion
along the line of sight of the emitting atmosphere. As the beam size
gets larger, more emission from the disk is included, and the effects
of self-absorption begin to be washed out. The model with 0.1
depletion factor is qualitatively very similar to the 0.5 depletion
factor case. There are minor differences involving the emission of
different velocity components in the proto-GGP, but generally the
results are the same. It is notable, however, that the peak line
temperature has dropped by a factor of 2-3. Self-absorption features
become more pronounced in the model with 0.01 depletion where an
inverse p-cygni profile is observed. In this lowest abundance model,
the column of emitting molecules is low enough that the emission can
no longer fill in the absorption trough. The absorption is washed out
for angular resolutions $\ga$0.1$\arcsec$. The emission from the rest
of the disk, once folded into the beam, quickly counteracts the
self-absorption feature and a single peaked Gaussian line is observed.

At all abundances in our models, the self absorption through the
proto-GGP reduces the flux at line center to 0 or negative
values. Conversely, non self-gravitating dense clumps in our models
only show moderate self-absorption. It may be that complete
self-absorption to zero or negative flux is a signature of GGP
formation discernible by ALMA. As the disk becomes more inclined, the
self-absorption features become less discernible as the emission line
broadens.

\section{Spectral Maps}

In Figure \ref{figure:1e-9specmap}, we have taken spectra along
various lines of sight through the face on protoplanetary disk with
0.5 depletion, convolved to a 0.007$\arcsec$ ALMA beam. The proto-GGP
stands out in emission quite high above the rest of the disk. This was
evident as well in the contour emission map (Figure
\ref{figure:map1}).

Spectral mosaics are particularly useful for identifying GGPs at low
HCO$^+$ abundance levels (e.g. Figure \ref{figure:5e-10specmap}) where
the proto-GGP may make itself evident through its bright emission line
and self-absorbed profile. This type of emission is missed in a
velocity integrated intensity contour map, however, owing to
self-absorption in the line.

\section{Conclusions}
In this paper, we have presented non-LTE radiative transfer
calculations of HCO$^+$ rotational line emission in a gravitationally
unstable protoplanetary disk. Our models suggest the following:
\begin{enumerate}
\item Dense gas clumps associated with proto-GGPs formed via disk
instabilities may be observable with interferometers (e.g. ALMA) in
molecular transitions with high critical density (e.g. HCO$^+$
J=7-6). The emission is dominated by radiatively pumped gas in the
outer layers of dense clumps, and thus does not necessarily reflect
the physical temperature of the emitting gas.

\item The emission lines arising from the densest self-gravitating gas
clumps in the protoplanetary disk are completely self-absorbed to zero
or negative flux. The line profiles from other dense gas clumps are
self absorbed, as well, although to a lesser degree. 

\item Proto-GGPs often appear as bright peaks in spectral line maps,
making them easier to identify in the presence of extended disk
emission.

\item ALMA will have the angular resolution and sensitivity necessary
to directly image proto-GGPs in formation in nearby circumstellar
disks.

\end{enumerate}

More complete chemical models and reaction networks will be
incorporated in our models in future works.

\acknowledgements We would like thank Adam Burrows and Philip Pinto
for enlightening conversations on radiative transfer and Monte Carlo
methods during code development.  Helpful conversations with Casey
Meakin during parallelization of our code are greatly appreciated as
well. We are grateful toward Michiel Hogerheijde and G-J van Zadelhoff
for providing the test results from the Leiden non-LTE radiative
transfer conference in digital format. We would additionally like to
thank Gopal Narayanan for his assistance with development of the line
of sight radiative transfer codes. We thank the referee for helpful
comments which have improved this work. D.N. acknowledges financial
support from an NSF Graduate Research Fellowship during this
study. The radiative transfer calculations were performed on the
Steward Observatory Beowulf cluster systems.

\clearpage

 \end{document}